\documentclass[aps,prl,twocolumn,superscriptaddress,showpacs]{revtex4-2}

\usepackage{color}
\usepackage{pstool}
\usepackage{amssymb}
\usepackage{units}
\usepackage{graphicx}
\usepackage{amsmath}
\usepackage{bm}

\newcommand{\rmd}{\mathrm{d}}

\newcommand{\lambdap}{\lambda_{\text{p}}}
\newcommand{\lambdaq}{\lambda_{\text{q}}}
\newcommand{\lambdar}{\lambda_{\text{r}}}
\newcommand{\lambdas}{\lambda_{\text{s}}}
\newcommand{\omegap}{\omega_{\text{p}}}
\newcommand{\omegaq}{\omega_{\text{q}}}
\newcommand{\omegar}{\omega_{\text{r}}}
\newcommand{\omegas}{\omega_{\text{s}}}

\newcommand{\taus}{\tau_{\text{s}}}

\definecolor{matlabgreen}{rgb}{0,0.5,0}
\EndPreamble

\begin{document}

\title{Toward a Quantum Memory in a Fiber Cavity Controlled by Intracavity Frequency Translation}
\author{Philip~J. Bustard}
 \email{philip.bustard@nrc-cnrc.gc.ca}
\affiliation{National Research Council of Canada, 100 Sussex Drive, Ottawa, Ontario, K1A 0R6, Canada}
\author{Kent Bonsma-Fisher}
\affiliation{National Research Council of Canada, 100 Sussex Drive, Ottawa, Ontario, K1A 0R6, Canada}
\author{Cyril Hnatovsky}
\affiliation{National Research Council of Canada, 100 Sussex Drive, Ottawa, Ontario, K1A 0R6, Canada}
\author{Dan Grobnic}
\affiliation{National Research Council of Canada, 100 Sussex Drive, Ottawa, Ontario, K1A 0R6, Canada}
\author{Stephen J. Mihailov}
\affiliation{National Research Council of Canada, 100 Sussex Drive, Ottawa, Ontario, K1A 0R6, Canada}
\author{Duncan England}
 \email{duncan.england@nrc-cnrc.gc.ca}
\affiliation{National Research Council of Canada, 100 Sussex Drive, Ottawa, Ontario, K1A 0R6, Canada}
\author{Benjamin~J. Sussman}
\affiliation{National Research Council of Canada, 100 Sussex Drive, Ottawa, Ontario, K1A 0R6, Canada}
\affiliation{Department of Physics, University of Ottawa, Ottawa, Ontario, K1N 6N5, Canada}
\date{\today}

\begin{abstract}
We propose a quantum memory protocol based on trapping photons in a fiber-integrated cavity, comprised of a birefringent fiber with dichroic reflective end facets. Photons are switched into resonance with the fiber cavity by intracavity Bragg-scattering frequency translation, driven by ancillary control pulses. After the storage delay, photons are switched out of resonance with the cavity, again by intracavity frequency translation. We demonstrate storage of quantum-level THz-bandwidth coherent states for a lifetime up to 16 cavity round trips, or \unit[200]{ns}, and a maximum overall efficiency of 73\%.
\end{abstract}

\maketitle

The ability to store information is a key capability for modern technologies, none more so than quantum technologies including communications~\cite{RevModPhys.74.145,Duan2001} and computing~\cite{Nature.409.46,Raussendorf2003}. Quantum memories capable of temporarily storing single photons, and other photonic states, will be critical for such technologies because photons are the natural carriers of quantum information, due to their high speeds and weak interactions. These attributes present a challenge: efficient, long-lived, and low-noise photon storage is difficult to implement. Research has yielded significant achievements in a wide variety of systems, using a range of protocols~\cite{Lvovsky2009,Simon2010,lasphotonrev4.244,JMO.63.S42}. The diversity of systems and protocols investigated is perhaps indicative of the difficulty in satisfying various, and often competing, criteria for practical use. This research multiplicity is likely to prove beneficial, however, since no single memory will satisfy all criteria such as efficiency, lifetime, wavelength, bandwidth, and mode capacity, for all applications.

A recurring theme in many memory designs is the idea of mapping a flying qubit (photon) into a stationary excitation in atomic vapor~\cite{PhysRevLett.86.783} or rare-earth ion-doped solid~\cite{nature.456.07607}. Such systems offer exceptional storage times~\cite{Katz2018} and may achieve the long-standing goal of repeaters for long-distance quantum communication~\cite{Duan2001,RevModPhys.83.33}. High-bandwidth designs~\cite{Reim2011} raise the prospect of broadband local photonic processing, particularly with low-noise `ladder' schemes~\cite{Kaczmarek2018,Finkelstein2018}.

An alternative memory design is to trap the target photon in a low-loss cavity~\cite{Pittman2002}. While the potential lifetimes are shorter than for the stationary excitation paradigm, relatively high time-bandwidth products may still be achieved with cavity storage, making such memories useful for temporal multiplexing~\cite{Pittman2002,Migdall2002}, for example. Successful temporal multiplexing of probabilistic photon pair sources~\cite{Kaneda2015,Kaneda2019,Xiong2016,Makino2016} might permit upscaling of multiphoton technologies~\cite{PhysRevLett.110.133601}, such as linear optical quantum computing~\cite{Nature.409.46} and measurement-based quantum computing~\cite{Raussendorf2003}, beyond the current state of the art~\cite{ZhongScience2020}. 

Fiber integration of storage cavities has been explored theoretically, both for a classical case with nonlinear optical switching of light~\cite{Sterke2002} and for quantum states with fast switching of cavity mirrors~\cite{Leung2006}. Such a memory could exploit fiber properties including: low-loss transmission; single-mode capacity; tight optical confinement; optimized commercial components; and, flexible routing. Indeed, fiber compatibility is often cited as a factor when designing other photonic components. A fiber-integrated cavity requires fast, low-noise switching, combined with sufficiently low loss to allow use of multiple independent bins, or cavity `round trips'. Despite the attractiveness of electrical switching~\cite{Margulis2011}, the insertion losses of fiber switches are a limiting factor. Light-by-light switching using $\chi^{(3)}$-nonlinearity in fiber offers an alternative solution that is fast and, theoretically, noiseless~\cite{OptExpress.13.9131}. Experiments have demonstrated high-bandwidth, high-efficiency, low-noise switching of single photons in fiber using the optical Kerr effect~\cite{Kupchak2019,Bouchard2021} and Bragg-scattering four-wave mixing (BSFWM)~\cite{McGuinness2010,OptLett.38.947,Clemmen2018}. 

\begin{figure*}
\begin{center}
\includegraphics[scale=1]{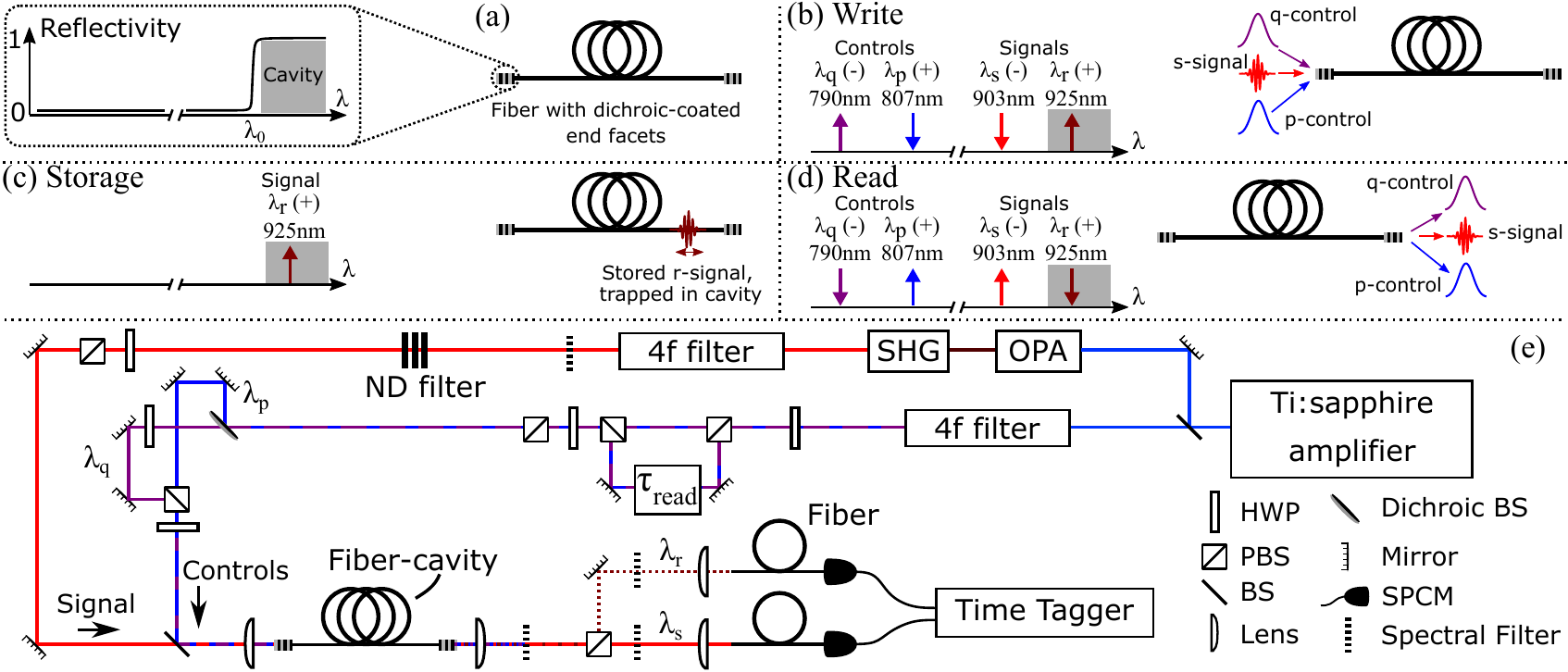}
\end{center}
\caption{(a) Fiber cavity, with dichroic-coated end facets. The fiber forms a cavity for wavelengths $\lambda>\lambda_0$ (shaded, gray). (b) Write-step: control pulses at $\lambdap$ on the slow axis (+) and $\lambdaq$ on the fast axis (-) redshift the signal photon from $\lambdas$ to $\lambdar$. (c) Storage: the signal photon at $\lambdar>\lambda_0$ is trapped in the cavity. (d) Read-step: control pulses at wavelengths $\lambdap$ and $\lambdaq$ blueshift the trapped signal photon from $\lambdar$ to $\lambdas$ so that it exits the fiber. (e) Experiment layout. See text for details. \label{fig:scheme}}
\end{figure*}

Here we propose and demonstrate an optical memory based on fiber-cavity storage with intracavity frequency translation (FC SWIFT). The fiber is a linear cavity, with the two end facets each coated with a short-wave-pass dichroic dielectric stack, such that for wavelengths $\lambda>\lambda_0$ the end facets have high reflectivity; for wavelengths $\lambda<\lambda_0$ light can enter and exit the fiber with high transmission (see Fig.~\ref{fig:scheme}(a)).
In the \textsl{write} step (Fig.~\ref{fig:scheme}(b)), the \textsl{signal} photon with wavelength $\lambdas<\lambda_0$ is input to the fiber with two intense ancillary \textsl{control} pulses at wavelengths $\lambdaq$ and $\lambdap$. The control pulses redshift the signal pulse from $\lambdas$ to $\lambdar>\lambda_0$ by BSFWM. As a result, the signal photon is reflected when it reaches the fiber end facet, and is thus trapped within the fiber cavity. During the storage stage (Fig.~\ref{fig:scheme}(c)), the signal photon continues to propagate back and forth within the fiber cavity. In the \textsl{read} step, control pulses at wavelengths $\lambdaq$ and $\lambdap$ propagate into the fiber timed to overlap with the circulating signal pulse. The signal photon is frequency-translated from $\lambdar$ to $\lambdas$ by BSFWM such that it exits the fiber. 

Cavities can be designed across a wide spectral range from the visible to the infrared, and broad bandwidths can be accommodated. In the case of a fiber cavity, integrated mirrors may minimize losses arising from multiple surface reflections in a free-space cavity and improve scalability with a smaller cavity footprint. The spatial mode of a fiber cavity will be well suited to operation with other composite fiber technologies. Indeed, integration of a fiber-based photon source~\cite{OptExpress.17.23589} within a FC SWIFT memory would reduce losses and equipment footprint, allowing efficient temporal multiplexing~\cite{Pittman2002,Migdall2002}.

\begin{figure}
\begin{center}
\includegraphics[trim=115 114 158 477,clip,width=\columnwidth]{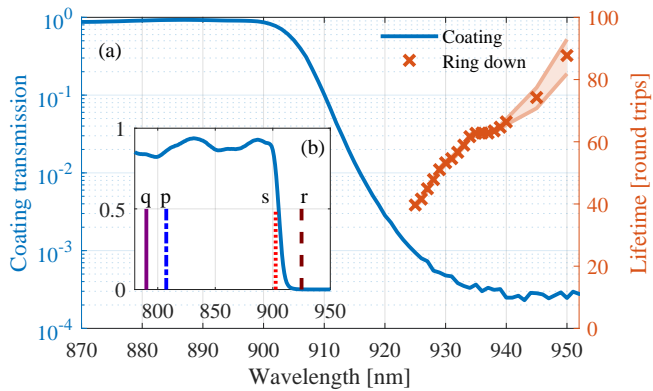}
\end{center}
\caption{Left axis: Coating transmission as a function of wavelength on (a) a log scale (blue curve) and (b) a linear scale (inset, blue curve). Wavelengths of the q-control (solid, purple), p-control (dash-dot, dark blue), s-signal (dots, red), and r-signal (dashed, maroon), are shown. Right axis: Measured cavity ring-down lifetime (red crosses) as a function of wavelength, with 95\% confidence intervals (shaded error bars). \label{fig:coatingVsWavelength}}
\end{figure}

We fabricated a fiber cavity in a polarization-maintaining (PM) fiber (\textsl{Fibercore} HB800). Birefringent PM fiber has been shown to offer the possibility of efficient, unidirectional frequency translation by BSFWM, with favorable phase-matching conditions for broadband signal pulses~\cite{OptExpress.26.17145}. Phase matching is achieved for unidirectional frequency translation with the control pulses launched on the two orthogonal polarization axes. In a fiber with normal dispersion, for appropriate control frequency separation the signal is frequency down-shifted by BSFWM when launched on the fast (-) axis, according to $\omegar=\omegas-(\omegaq-\omegap)$, where $\omega_i=\nicefrac{2\pi c}{\lambda_i}$~\cite{OptExpress.26.17145}. Figure~\ref{fig:coatingVsWavelength}(a) and inset (b) show the coating transmission $\mathcal{T}$ measured on a blank substrate during the coating run (\textsl{Omega Optical}). The coating was designed to yield high transmission for the control pulses at $\lambdaq=\unit[790.1]{nm}$ and $\lambdap=\unit[807.4]{nm}$, and for the input signal pulses at $\lambdas=\unit[902.5]{nm}$; at wavelengths $\lambda>\unit[920]{nm}$ the coating has high reflectivity to form a cavity, including the storage wavelength $\lambdar=\unit[925]{nm}$.

We measured the cavity lifetime as a function of wavelength using a ring-down procedure. Bright signal pulses (\unit[$\gtrsim10^5$]{photons/pulse}) with a full-width at half-maximum (FWHM) bandwidth of $\approx\unit[1]{nm}$ were coupled into the 1.285-m end-facet-coated (EFC) PM fiber, polarized along the slow axis. Light exiting the fiber at the output facet was coupled into a single-mode collection fiber and detected using a single-photon-counting module (SPCM). The time delay of each detection event relative to the laser trigger was recorded to build up a histogram of detection events. The resulting histograms show a comb structure of regular peaks spaced spaced by $\tau_{\text{rt}}=\unit[12.67]{ns}$ ($\times1$ cavity round trip) and decaying over several microseconds. Figure~\ref{fig:coatingVsWavelength} (right axis) shows a plot of the ring-down lifetime extracted using a weighted linear regression on the summed counts in each ``round trip'' bin.
The measured $1/\rm{e}$ lifetime at \unit[925]{nm} is $39.7(5)\tau_{\text{rt}}$, and this increases to $87(6)\tau_{\text{rt}}$ at \unit[950]{nm}, where the reduced count rate limited further wavelength tuning. As expected, the cavity lifetime increases with wavelength, since the reflectivity of the coating is also increasing. The cavity lifetimes are sufficiently long to offer promise for use as an FC SWIFT memory.

Figure~\ref{fig:scheme}(e) shows the experiment layout. The main laser is a Ti:sapphire amplifier operating  at $\mathcal{R}=\unit[1]{kHz}$; it outputs 80-fs pulses, with FWHM bandwidth $\Delta\lambda=\unit[12]{nm}$ centered at \unit[800]{nm}. The laser output is partitioned at a beam splitter to prepare signal and control pulses. In the control `arm', we use a 4$f$ spectral filter~\cite{Weiner1988} to prepare control pulses at $\lambdaq=\unit[790.1]{nm}$ and $\lambdap=\unit[807.4]{nm}$, each with FWHM $\Delta\lambda=\unit[0.4]{nm}$, corresponding to pulse durations of \unit[2.3]{ps}. We prepare distinct write and read pulse pairs using an imbalanced interferometer, with the read control pulses delayed by $\tau_{\text{read}}\approx\unit[12.67]{ns}$ relative to the write control pulses; the p- and q-controls are separated and recombined with orthogonal polarizations. In the signal arm, signal pulses are prepared by second harmonic generation (SHG) of the \unit[1.805]{$\mu$m} output from an optical parametric amplifier (OPA); the SHG light is spectrally filtered in a 4$f$ spectral filter. The signal pulses have a bandwidth of $\Delta\lambdas=\unit[1.05]{nm}$ centered at $\lambdas=\unit[902.5]{nm}$, corresponding to a pulse duration of $\Delta\taus=\unit[1.1]{ps}$. These input wavelengths give a storage wavelength of $\lambdar=\unit[925]{nm}$. The control pulses are combined with the signal pulse at a beam splitter and coupled into the fiber cavity. The coating-corrected coupling efficiencies for the write and read controls were $\eta_{\text{q}}^{\text{w}}=0.50(3)$, $\eta_{\text{p}}^{\text{w}}=0.41(3)$ and $\eta_{\text{q}}^{\text{r}}=0.44(3)$, $\eta_{\text{p}}^{\text{r}}=0.38(3)$, respectively. Control pulse energies were set to deliver $W^{\text{out}}_{\text{q}}=W^{\text{out}}_{\text{p}}=\unit[2.2(3)]{nJ}$ at the fiber exit facet. At the fiber-cavity exit, signal photons polarized on the fast and slow axes are coupled to separate single mode fibers (SMFs). Interference filters are arranged prior to the fast- and slow-axis collection fibers to give transmission windows of $\approx\unit[3]{nm}$ bandwidth at \unit[902.5]{nm} and \unit[925]{nm}, respectively. The collection fibers are each coupled to SPCMs, and detection events are recorded. We performed experiments with input signal pulses attenuated to the single-photon level (SPL), where memory noise properties can be measured. When count rates were insufficient to measure the spectrum, or photons leaking from the cavity, we used ``bright'' input signal pulses (\unit[$\sim10^5$]{photons/pulse}), where the memory operation remains linear in the signal power.

\begin{figure}
\begin{center}
\includegraphics[trim=115 115 158 468,clip,width=\columnwidth]{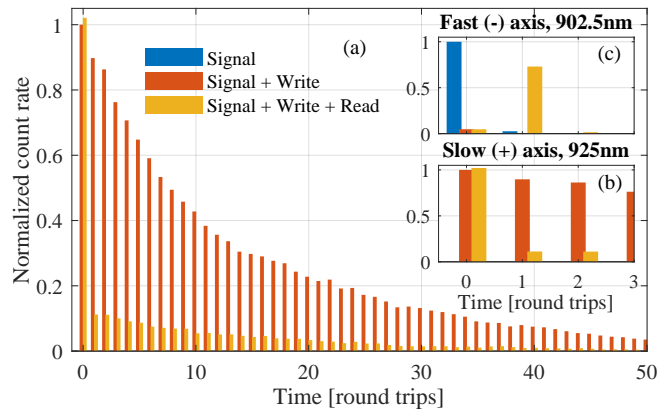}
\end{center}
\caption{Histograms of the counts, normalized to the rate at $T=0$, measured on (a,b) the slow axis and (c) the fast axis for a bright signal pulse input on the fast axis with: no control pulses (blue); write controls on (red); and, write and read controls on (yellow). (a) Ring-down showing storage of the signal pulse circulating in the cavity. The shaded region corresponds to the interval plotted in the insets. (b) Read-out at $T=1$ reduces the number of photons circulating in the cavity on the slow axis. (c) Storage at $T=0$ and read-out at $T=1$ (\unit[12.67]{ns} delay).  \label{fig:brightSignalRingDown}}
\end{figure}

We first characterize the memory operation using bright signal pulses input with polarization along the fiber fast axis; we insert neutral density filters before the collection fibers to avoid saturating the SPCMs. Figure~\ref{fig:brightSignalRingDown} plots the normalized SPCM count rate on (a,b) the slow axis and (c) the fast axis, after time tagging and binning the detection events. The count rates are plotted as a function of round trip number $T=t/\tau_{\text{rt}}$, where $t$ is time.
First, we consider the write step, comparing the results with the write controls off (blue bars) and on (red bars). For optimal delay between the signal and write p-, q-control pulses, the fast axis count rate at $T=0$ is diminished with write efficiency $\eta_{\text{w}}=0.95(2)$ (see Fig.~\ref{fig:brightSignalRingDown}(c)). On the slow axis, a clear ring down signal of \unit[925]{nm} light leaking through the cavity exit facet is measured (red bars), with a $1/\text{e}$ lifetime of approximately $\tau_{\text{c}}\approx16\tau_{\text{rt}}\approx\unit[206]{ns}$: these signal photons have been redshifted and polarization-switched on to the slow axis such that they are trapped within the cavity, before leaking out.  Applying the read control pulses at a write-read delay of $\unit[12.67]{ns}$, corresponding to one round trip of the cavity, we observe that the slow axis count rate at $T=1$ reduces by a factor $\eta_{\text{r}}=0.87(4)$ and the count rate on the fast axis increases, showing that signal photons have been blueshifted and polarization-switched back to the fast axis, such that they are released from the cavity. Comparing the fast axis count rates at $T=1$ with all controls on, to the rates at $T=0$ with the signal only, we measure a total efficiency of $\eta_{\text{tot}}=0.73(1)$.

We attenuated the signal to achieve $\mathcal{N_{\text{in}}}=\unit[1.0(1)]{photons/pulse}$ inside the fiber at wavelength $\lambdas$~\footnote{See Supplemental Material for experimental details}. Figure~\ref{fig:readDelayScanSPL} (left axis) shows a plot of the signal, noise, and corrected signal as a function of (a) signal delay relative to fixed write control pulses, and (b) read controls' delay relative to fixed signal and write control pulses. The background-corrected efficiency is $\eta^{\text{SPL}}_{\text{tot}}=0.70(4)$, consistent with the measurement using bright signal pulses. The temporal profile of the delay scans is determined by the pulse durations and the group velocity walk-off between the signal and control pulses~\cite{OptExpress.26.17145}. For the known pulse durations and analytic solutions~\cite{OptExpress.26.17145}, we extract a temporal walk-off of $\Delta_{\text{w}}=\unit[16.6]{ps}$; this is consistent with the expected walk-off of \unit[17]{ps}, estimated using dispersion data. The maximum signal-to-noise ratio (SNR) is $\mathcal{S}=2.3(1)$, with noise level $\mathcal{N_{\text{noise}}}=\unit[0.30(3)]{photons/pulse}$. This gives a noise performance benchmark $\mu_1=\mathcal{N_{\text{noise}}}/\eta^{\text{SPL}}_{\text{tot}}=0.43(5)$. The noise due to the write and read control pulses alone (Fig.~\ref{fig:readDelayScanSPL}(b)) is delay dependent in the readout bin, reaching a maximum near the peak of the retrieved signal, indicating that the read control pulses are reading out noise photons created by the write control pulses.
\begin{figure}
\begin{center}
\includegraphics[trim=115 128 158 470,clip,width=\columnwidth]{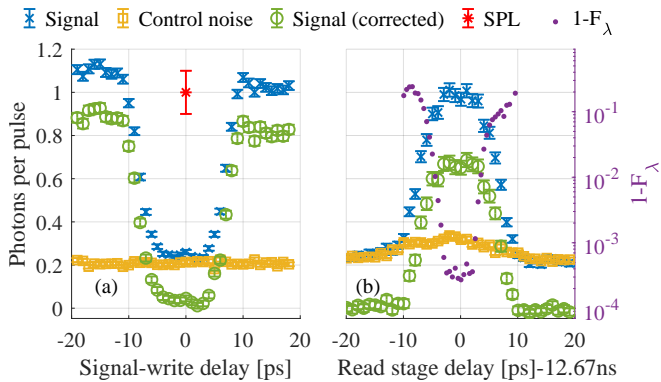}
\end{center}
\caption{Left ordinate: Fast axis detection rate for an input signal at the single-photon level (red, asterisk) as a function (a) of signal delay relative to fixed write control pulses, and (b) of read stage delay relative to fixed write and signal stage delays. Plotted are the signal (blue crosses), noise (yellow squares), and background-corrected signal (green circles). Poisson error bars are shown for each case. Right ordinate (log-scale): Deviation of the classical spectral fidelity $F_\lambda$ from 1, as a function of the read stage delay (purple dots). \label{fig:readDelayScanSPL}}
\end{figure}

For a quantitative comparison of the memory operation, using bright signal pulses we measure the classical spectral fidelity $F_{\lambda}=\left.\int\sqrt{I_{\text{in}}(\lambda)I_{\text{out}}(\lambda)} \rmd\lambda\, \middle/\left[\int I_{\text{in}}(\lambda)\rmd\lambda \int I_{\text{out}}(\lambda)\rmd\lambda\right]^{1/2}\right.$ between the signal spectral intensity without ($I_{\text{in}}$) and with ($I_{\text{out}}$) the memory interaction, assuming constant spectral phase. An ideal memory will have $F_{\lambda}=1$, such that the input and output spectra are identical apart from the group delay introduced by the memory. Figure~\ref{fig:readDelayScanSPL}(b) plots $1-F_{\lambda}$ (purple dots, right axis) as a function of the read stage delay, showing maximum spectral fidelity of $F_{\text{max}}>99.97\%$ at the optimum delay. Comparing the fidelity plot with the SPL readout plot, we see that the read efficiency and spectral fidelity are both reduced when the read delays are such that the signal pulse only completes a partial ``collision'' with the Bragg grating created by the control pulses. At these delay settings, the effect of cross-phase modulation from the controls is not symmetric on the signal, such that the signal is partially redshifted or blueshifted relative to the input spectrum, and the spectral fidelity diminishes.

We now proceed with a discussion of the results. For end-facet coatings with unit reflectivity, the maximum lifetime possible at the current cavity length is $\sim338\tau_{\text{rt}}$, given the quoted fiber loss of $\sim\unit[5]{dB/km}$. The lifetime measured in Fig.~\ref{fig:brightSignalRingDown} is $\tau_{\text{c}}\approx16\tau_{\text{rt}}$, whereas the lifetime when measured with a resonant signal probe at $\lambdar=\unit[925]{nm}$ was $\tau_{\text{c}}=39.7(4)\tau_{\text{rt}}$; the lifetime at $\unit[950]{nm}$ was $\tau_{\text{c}}=87(6)\tau_{\text{rt}}$. At input pulse energies of $W_{\text{in}}\approx\unit[8-10]{nJ}$, we observed irreparable damage to EFC fibers and therefore restricted our operating range to $W_{\text{in}}<\unit[7.5]{nJ}$. We attribute the decrease in the lifetime to coating damage caused by the control pulses. We expect that better mode matching and improved coating designs will reduce the power requirements, enhancing the cavity lifetime and durability. The pulse energy restrictions also limited the total efficiency by reducing the BSFWM interaction strength. While we were not able to achieve unit frequency translation efficiency in either write or read steps, we were able to measure translation efficiencies as high as 98\% in an uncoated fiber, suggesting that higher memory efficiencies will be possible with design improvements to increase the control pulse energies within the fiber cavity.

The maximum SNR demonstrated at the SPL was $\mathcal{S}=2.3(1)$, with spontaneous Raman scattering from the control pulses as the main noise source~\footnote{See Supplemental Material for noise measurements.}. This performance is insufficient for practical use; nonetheless, improvements are feasible. The Raman gain of silica diminishes significantly for shifts $\gtrsim$\unit[40]{THz}~\cite{JOSAB.6.1159}: noise should reduce with increased signal-control detuning. Alternatively, if signal frequencies were on the anti-Stokes side of the controls, the Raman noise would be further reduced.

A memory must be able to store and retrieve photons ``on demand'', or at least in synchronization with an appropriate clock rate. To accommodate the inherent periodicity of the FC SWIFT protocol, a modelocked laser, or lasers, can be locked to the fiber-cavity frequency, allowing synchronous pumping of photon pair sources and memory operation. Combined with cavity fabrication improvements, use of synchronous control lasers should enable retrieval in $\tau_\text{c}/\tau_\text{rt}\sim100$ distinct time-bins. While this is modest compared to the highest time-bandwidth products, here the cavity round-trip time is much longer than the signal  duration, $\tau_{\text{rt}}/\Delta\tau_{\text{s}}\approx1.1\times10^4$, such that feed-forward control may be feasible.

A practical memory must also achieve high fidelity between the input and retrieved states. Beyond the influence of noise and loss, any distortions introduced by the memory interaction will reduce the fidelity. Our measurement of the classical spectral fidelity indicates that the spectrum is not significantly distorted by the two frequency conversions. Nonetheless, for longer storage times, dispersion will play an important role; for this reason, the FC SWIFT memory may be most appropriate for telecom wavelengths where dispersion and loss are lowest. It may also be possible to compensate for fiber dispersion with a chirped dichroic end-facet coating. 

We stored THz-bandwidth SPL coherent states in an EFC birefringent fiber, using BSFWM to switch photons in and out of resonance with the dichroic reflective end facets, at a high efficiency. We hope that the FC SWIFT memory will be useful for multiplexing of probabilistic photon sources~\cite{OptExpress.17.23589}.

\begin{acknowledgements}
The authors are grateful for discussions with Khabat Heshami, Fr\'ed\'eric Bouchard, Yingwen Zhang, Kate Fenwick, Rune Lausten, Doug Moffatt, and Denis Guay. P.J.B. thanks Gary Carver for the use of transmission data.
\end{acknowledgements}


\begin{thebibliography}{38}%
\makeatletter
\providecommand \@ifxundefined [1]{%
 \@ifx{#1\undefined}
}%
\providecommand \@ifnum [1]{%
 \ifnum #1\expandafter \@firstoftwo
 \else \expandafter \@secondoftwo
 \fi
}%
\providecommand \@ifx [1]{%
 \ifx #1\expandafter \@firstoftwo
 \else \expandafter \@secondoftwo
 \fi
}%
\providecommand \natexlab [1]{#1}%
\providecommand \enquote  [1]{``#1''}%
\providecommand \bibnamefont  [1]{#1}%
\providecommand \bibfnamefont [1]{#1}%
\providecommand \citenamefont [1]{#1}%
\providecommand \href@noop [0]{\@secondoftwo}%
\providecommand \href [0]{\begingroup \@sanitize@url \@href}%
\providecommand \@href[1]{\@@startlink{#1}\@@href}%
\providecommand \@@href[1]{\endgroup#1\@@endlink}%
\providecommand \@sanitize@url [0]{\catcode `\\12\catcode `\$12\catcode
  `\&12\catcode `\#12\catcode `\^12\catcode `\_12\catcode `\%12\relax}%
\providecommand \@@startlink[1]{}%
\providecommand \@@endlink[0]{}%
\providecommand \url  [0]{\begingroup\@sanitize@url \@url }%
\providecommand \@url [1]{\endgroup\@href {#1}{\urlprefix }}%
\providecommand \urlprefix  [0]{URL }%
\providecommand \Eprint [0]{\href }%
\providecommand \doibase [0]{https://doi.org/}%
\providecommand \selectlanguage [0]{\@gobble}%
\providecommand \bibinfo  [0]{\@secondoftwo}%
\providecommand \bibfield  [0]{\@secondoftwo}%
\providecommand \translation [1]{[#1]}%
\providecommand \BibitemOpen [0]{}%
\providecommand \bibitemStop [0]{}%
\providecommand \bibitemNoStop [0]{.\EOS\space}%
\providecommand \EOS [0]{\spacefactor3000\relax}%
\providecommand \BibitemShut  [1]{\csname bibitem#1\endcsname}%
\let\auto@bib@innerbib\@empty
\bibitem [{\citenamefont {Gisin}\ \emph {et~al.}(2002)\citenamefont {Gisin},
  \citenamefont {Ribordy}, \citenamefont {Tittel},\ and\ \citenamefont
  {Zbinden}}]{RevModPhys.74.145}%
  \BibitemOpen
  \bibfield  {author} {\bibinfo {author} {\bibfnamefont {N.}~\bibnamefont
  {Gisin}}, \bibinfo {author} {\bibfnamefont {G.}~\bibnamefont {Ribordy}},
  \bibinfo {author} {\bibfnamefont {W.}~\bibnamefont {Tittel}},\ and\ \bibinfo
  {author} {\bibfnamefont {H.}~\bibnamefont {Zbinden}},\ }\bibfield  {title}
  {\bibinfo {title} {Quantum cryptography},\ }\href
  {https://doi.org/10.1103/RevModPhys.74.145} {\bibfield  {journal} {\bibinfo
  {journal} {Rev. Mod. Phys.}\ }\textbf {\bibinfo {volume} {74}},\ \bibinfo
  {pages} {145} (\bibinfo {year} {2002})}\BibitemShut {NoStop}%
\bibitem [{\citenamefont {Duan}\ \emph {et~al.}(2001)\citenamefont {Duan},
  \citenamefont {Lukin}, \citenamefont {Cirac},\ and\ \citenamefont
  {Zoller}}]{Duan2001}%
  \BibitemOpen
  \bibfield  {author} {\bibinfo {author} {\bibfnamefont {L.-M.}\ \bibnamefont
  {Duan}}, \bibinfo {author} {\bibfnamefont {M.~D.}\ \bibnamefont {Lukin}},
  \bibinfo {author} {\bibfnamefont {J.~I.}\ \bibnamefont {Cirac}},\ and\
  \bibinfo {author} {\bibfnamefont {P.}~\bibnamefont {Zoller}},\ }\bibfield
  {title} {\bibinfo {title} {Long-distance quantum communication with atomic
  ensembles and linear optics},\ }\href@noop {} {\bibfield  {journal} {\bibinfo
   {journal} {Nature}\ }\textbf {\bibinfo {volume} {414}},\ \bibinfo {pages}
  {413} (\bibinfo {year} {2001})}\BibitemShut {NoStop}%
\bibitem [{\citenamefont {Knill}\ \emph {et~al.}(2001)\citenamefont {Knill},
  \citenamefont {Laflamme},\ and\ \citenamefont {Milburn}}]{Nature.409.46}%
  \BibitemOpen
  \bibfield  {author} {\bibinfo {author} {\bibfnamefont {E.}~\bibnamefont
  {Knill}}, \bibinfo {author} {\bibfnamefont {R.}~\bibnamefont {Laflamme}},\
  and\ \bibinfo {author} {\bibfnamefont {G.~J.}\ \bibnamefont {Milburn}},\
  }\bibfield  {title} {\bibinfo {title} {A scheme for efficient quantum
  computation with linear optics},\ }\href@noop {} {\bibfield  {journal}
  {\bibinfo  {journal} {Nature}\ }\textbf {\bibinfo {volume} {409}},\ \bibinfo
  {pages} {46} (\bibinfo {year} {2001})}\BibitemShut {NoStop}%
\bibitem [{\citenamefont {Raussendorf}\ \emph {et~al.}(2003)\citenamefont
  {Raussendorf}, \citenamefont {Browne},\ and\ \citenamefont
  {Briegel}}]{Raussendorf2003}%
  \BibitemOpen
  \bibfield  {author} {\bibinfo {author} {\bibfnamefont {R.}~\bibnamefont
  {Raussendorf}}, \bibinfo {author} {\bibfnamefont {D.~E.}\ \bibnamefont
  {Browne}},\ and\ \bibinfo {author} {\bibfnamefont {H.~J.}\ \bibnamefont
  {Briegel}},\ }\bibfield  {title} {\bibinfo {title} {Measurement-based quantum
  computation on cluster states},\ }\href
  {https://doi.org/10.1103/physreva.68.022312} {\bibfield  {journal} {\bibinfo
  {journal} {Phys. Rev. A}\ }\textbf {\bibinfo {volume} {68}},\ \bibinfo
  {pages} {022312} (\bibinfo {year} {2003})}\BibitemShut {NoStop}%
\bibitem [{\citenamefont {Lvovsky}\ \emph {et~al.}(2009)\citenamefont
  {Lvovsky}, \citenamefont {Sanders},\ and\ \citenamefont
  {Tittel}}]{Lvovsky2009}%
  \BibitemOpen
  \bibfield  {author} {\bibinfo {author} {\bibfnamefont {A.}~\bibnamefont
  {Lvovsky}}, \bibinfo {author} {\bibfnamefont {B.}~\bibnamefont {Sanders}},\
  and\ \bibinfo {author} {\bibfnamefont {W.}~\bibnamefont {Tittel}},\
  }\bibfield  {title} {\bibinfo {title} {Optical quantum memory},\ }\href@noop
  {} {\bibfield  {journal} {\bibinfo  {journal} {Nat. Photon.}\ }\textbf
  {\bibinfo {volume} {3}},\ \bibinfo {pages} {706} (\bibinfo {year}
  {2009})}\BibitemShut {NoStop}%
\bibitem [{\citenamefont {Simon}\ \emph {et~al.}(2010)\citenamefont {Simon},
  \citenamefont {Afzelius}, \citenamefont {Appel}, \citenamefont {{Boyer de la
  Giroday}}, \citenamefont {Dewhurst}, \citenamefont {Gisin}, \citenamefont
  {Hu}, \citenamefont {Jelezko}, \citenamefont {Kr{\"{o}}ll}, \citenamefont
  {M{\"{u}}ller}, \citenamefont {Nunn}, \citenamefont {Polzik}, \citenamefont
  {Rarity}, \citenamefont {De~Riedmatten}, \citenamefont {Rosenfeld},
  \citenamefont {Shields}, \citenamefont {Sk{\"{o}}ld}, \citenamefont
  {Stevenson}, \citenamefont {Thew}, \citenamefont {Walmsley}, \citenamefont
  {Weber}, \citenamefont {Weinfurter}, \citenamefont {Wrachtrup},\ and\
  \citenamefont {Young}}]{Simon2010}%
  \BibitemOpen
  \bibfield  {author} {\bibinfo {author} {\bibfnamefont {C.}~\bibnamefont
  {Simon}}, \bibinfo {author} {\bibfnamefont {M.}~\bibnamefont {Afzelius}},
  \bibinfo {author} {\bibfnamefont {J.}~\bibnamefont {Appel}}, \bibinfo
  {author} {\bibfnamefont {A.}~\bibnamefont {{Boyer de la Giroday}}}, \bibinfo
  {author} {\bibfnamefont {S.~J.}\ \bibnamefont {Dewhurst}}, \bibinfo {author}
  {\bibfnamefont {N.}~\bibnamefont {Gisin}}, \bibinfo {author} {\bibfnamefont
  {C.~Y.}\ \bibnamefont {Hu}}, \bibinfo {author} {\bibfnamefont
  {F.}~\bibnamefont {Jelezko}}, \bibinfo {author} {\bibfnamefont
  {S.}~\bibnamefont {Kr{\"{o}}ll}}, \bibinfo {author} {\bibfnamefont {J.~H.}\
  \bibnamefont {M{\"{u}}ller}}, \bibinfo {author} {\bibfnamefont
  {J.}~\bibnamefont {Nunn}}, \bibinfo {author} {\bibfnamefont {E.~S.}\
  \bibnamefont {Polzik}}, \bibinfo {author} {\bibfnamefont {J.~G.}\
  \bibnamefont {Rarity}}, \bibinfo {author} {\bibfnamefont {H.}~\bibnamefont
  {De~Riedmatten}}, \bibinfo {author} {\bibfnamefont {W.}~\bibnamefont
  {Rosenfeld}}, \bibinfo {author} {\bibfnamefont {A.~J.}\ \bibnamefont
  {Shields}}, \bibinfo {author} {\bibfnamefont {N.}~\bibnamefont
  {Sk{\"{o}}ld}}, \bibinfo {author} {\bibfnamefont {R.~M.}\ \bibnamefont
  {Stevenson}}, \bibinfo {author} {\bibfnamefont {R.}~\bibnamefont {Thew}},
  \bibinfo {author} {\bibfnamefont {I.~A.}\ \bibnamefont {Walmsley}}, \bibinfo
  {author} {\bibfnamefont {M.~C.}\ \bibnamefont {Weber}}, \bibinfo {author}
  {\bibfnamefont {H.}~\bibnamefont {Weinfurter}}, \bibinfo {author}
  {\bibfnamefont {J.}~\bibnamefont {Wrachtrup}},\ and\ \bibinfo {author}
  {\bibfnamefont {R.~J.}\ \bibnamefont {Young}},\ }\bibfield  {title} {\bibinfo
  {title} {Quantum memories},\ }\href
  {http://dx.doi.org/10.1140/epjd/e2010-00103-y} {\bibfield  {journal}
  {\bibinfo  {journal} {Eur. Phys. J. D}\ }\textbf {\bibinfo {volume} {58}},\
  \bibinfo {pages} {1} (\bibinfo {year} {2010})}\BibitemShut {NoStop}%
\bibitem [{\citenamefont {Tittel}\ \emph {et~al.}(2010)\citenamefont {Tittel},
  \citenamefont {Afzelius}, \citenamefont {Chaneli{\'e}re}, \citenamefont
  {Cone}, \citenamefont {Kr{\"{o}}ll}, \citenamefont {Moiseev},\ and\
  \citenamefont {Sellars}}]{lasphotonrev4.244}%
  \BibitemOpen
  \bibfield  {author} {\bibinfo {author} {\bibfnamefont {W.}~\bibnamefont
  {Tittel}}, \bibinfo {author} {\bibfnamefont {M.}~\bibnamefont {Afzelius}},
  \bibinfo {author} {\bibfnamefont {T.}~\bibnamefont {Chaneli{\'e}re}},
  \bibinfo {author} {\bibfnamefont {R.}~\bibnamefont {Cone}}, \bibinfo {author}
  {\bibfnamefont {S.}~\bibnamefont {Kr{\"{o}}ll}}, \bibinfo {author}
  {\bibfnamefont {S.}~\bibnamefont {Moiseev}},\ and\ \bibinfo {author}
  {\bibfnamefont {M.}~\bibnamefont {Sellars}},\ }\bibfield  {title} {\bibinfo
  {title} {Photon-echo quantum memory in solid state systems},\ }\href
  {https://doi.org/10.1002/lpor.200810056} {\bibfield  {journal} {\bibinfo
  {journal} {Laser Photonics Rev.}\ }\textbf {\bibinfo {volume} {4}},\ \bibinfo
  {pages} {244} (\bibinfo {year} {2010})}\BibitemShut {NoStop}%
\bibitem [{\citenamefont {Heshami}\ \emph {et~al.}(2016)\citenamefont
  {Heshami}, \citenamefont {England}, \citenamefont {Humphreys}, \citenamefont
  {Bustard}, \citenamefont {Acosta}, \citenamefont {Nunn},\ and\ \citenamefont
  {Sussman}}]{JMO.63.S42}%
  \BibitemOpen
  \bibfield  {author} {\bibinfo {author} {\bibfnamefont {K.}~\bibnamefont
  {Heshami}}, \bibinfo {author} {\bibfnamefont {D.~G.}\ \bibnamefont
  {England}}, \bibinfo {author} {\bibfnamefont {P.~C.}\ \bibnamefont
  {Humphreys}}, \bibinfo {author} {\bibfnamefont {P.~J.}\ \bibnamefont
  {Bustard}}, \bibinfo {author} {\bibfnamefont {V.~M.}\ \bibnamefont {Acosta}},
  \bibinfo {author} {\bibfnamefont {J.}~\bibnamefont {Nunn}},\ and\ \bibinfo
  {author} {\bibfnamefont {B.~J.}\ \bibnamefont {Sussman}},\ }\bibfield
  {title} {\bibinfo {title} {Quantum memories: emerging applications and recent
  advances},\ }\href {https://doi.org/10.1080/09500340.2016.1148212} {\bibfield
   {journal} {\bibinfo  {journal} {J. Mod. Opt.}\ }\textbf {\bibinfo {volume}
  {63}},\ \bibinfo {pages} {2005} (\bibinfo {year} {2016})}\BibitemShut
  {NoStop}%
\bibitem [{\citenamefont {Phillips}\ \emph {et~al.}(2001)\citenamefont
  {Phillips}, \citenamefont {Fleischhauer}, \citenamefont {Mair}, \citenamefont
  {Walsworth},\ and\ \citenamefont {Lukin}}]{PhysRevLett.86.783}%
  \BibitemOpen
  \bibfield  {author} {\bibinfo {author} {\bibfnamefont {D.~F.}\ \bibnamefont
  {Phillips}}, \bibinfo {author} {\bibfnamefont {A.}~\bibnamefont
  {Fleischhauer}}, \bibinfo {author} {\bibfnamefont {A.}~\bibnamefont {Mair}},
  \bibinfo {author} {\bibfnamefont {R.~L.}\ \bibnamefont {Walsworth}},\ and\
  \bibinfo {author} {\bibfnamefont {M.~D.}\ \bibnamefont {Lukin}},\ }\bibfield
  {title} {\bibinfo {title} {Storage of light in atomic vapor},\ }\href
  {https://doi.org/10.1103/PhysRevLett.86.783} {\bibfield  {journal} {\bibinfo
  {journal} {Phys. Rev. Lett.}\ }\textbf {\bibinfo {volume} {86}},\ \bibinfo
  {pages} {783} (\bibinfo {year} {2001})}\BibitemShut {NoStop}%
\bibitem [{\citenamefont {de~Riedmatten}\ \emph {et~al.}(2008)\citenamefont
  {de~Riedmatten}, \citenamefont {Afzelius}, \citenamefont {Staudt},
  \citenamefont {Simon},\ and\ \citenamefont {Gisin}}]{nature.456.07607}%
  \BibitemOpen
  \bibfield  {author} {\bibinfo {author} {\bibfnamefont {H.}~\bibnamefont
  {de~Riedmatten}}, \bibinfo {author} {\bibfnamefont {M.}~\bibnamefont
  {Afzelius}}, \bibinfo {author} {\bibfnamefont {M.}~\bibnamefont {Staudt}},
  \bibinfo {author} {\bibfnamefont {C.}~\bibnamefont {Simon}},\ and\ \bibinfo
  {author} {\bibfnamefont {N.}~\bibnamefont {Gisin}},\ }\bibfield  {title}
  {\bibinfo {title} {A solid-state light--matter interface at the single-photon
  level},\ }\href@noop {} {\bibfield  {journal} {\bibinfo  {journal} {Nature}\
  }\textbf {\bibinfo {volume} {456}},\ \bibinfo {pages} {773} (\bibinfo {year}
  {2008})}\BibitemShut {NoStop}%
\bibitem [{\citenamefont {Katz}\ and\ \citenamefont
  {Firstenberg}(2018)}]{Katz2018}%
  \BibitemOpen
  \bibfield  {author} {\bibinfo {author} {\bibfnamefont {O.}~\bibnamefont
  {Katz}}\ and\ \bibinfo {author} {\bibfnamefont {O.}~\bibnamefont
  {Firstenberg}},\ }\bibfield  {title} {\bibinfo {title} {Light storage for one
  second in room-temperature alkali vapor},\ }\href
  {https://doi.org/10.1038/s41467-018-04458-4} {\bibfield  {journal} {\bibinfo
  {journal} {Nat. Commun.}\ }\textbf {\bibinfo {volume} {9}},\ \bibinfo {pages}
  {2074} (\bibinfo {year} {2018})}\BibitemShut {NoStop}%
\bibitem [{\citenamefont {Sangouard}\ \emph {et~al.}(2011)\citenamefont
  {Sangouard}, \citenamefont {Simon}, \citenamefont {de~Riedmatten},\ and\
  \citenamefont {Gisin}}]{RevModPhys.83.33}%
  \BibitemOpen
  \bibfield  {author} {\bibinfo {author} {\bibfnamefont {N.}~\bibnamefont
  {Sangouard}}, \bibinfo {author} {\bibfnamefont {C.}~\bibnamefont {Simon}},
  \bibinfo {author} {\bibfnamefont {H.}~\bibnamefont {de~Riedmatten}},\ and\
  \bibinfo {author} {\bibfnamefont {N.}~\bibnamefont {Gisin}},\ }\bibfield
  {title} {\bibinfo {title} {Quantum repeaters based on atomic ensembles and
  linear optics},\ }\href {https://doi.org/10.1103/RevModPhys.83.33} {\bibfield
   {journal} {\bibinfo  {journal} {Rev. Mod. Phys.}\ }\textbf {\bibinfo
  {volume} {83}},\ \bibinfo {pages} {33} (\bibinfo {year} {2011})}\BibitemShut
  {NoStop}%
\bibitem [{\citenamefont {Reim}\ \emph {et~al.}(2011)\citenamefont {Reim},
  \citenamefont {Michelberger}, \citenamefont {Lee}, \citenamefont {Nunn},
  \citenamefont {Langford},\ and\ \citenamefont {Walmsley}}]{Reim2011}%
  \BibitemOpen
  \bibfield  {author} {\bibinfo {author} {\bibfnamefont {K.~F.}\ \bibnamefont
  {Reim}}, \bibinfo {author} {\bibfnamefont {P.}~\bibnamefont {Michelberger}},
  \bibinfo {author} {\bibfnamefont {K.~C.}\ \bibnamefont {Lee}}, \bibinfo
  {author} {\bibfnamefont {J.}~\bibnamefont {Nunn}}, \bibinfo {author}
  {\bibfnamefont {N.~K.}\ \bibnamefont {Langford}},\ and\ \bibinfo {author}
  {\bibfnamefont {I.~A.}\ \bibnamefont {Walmsley}},\ }\bibfield  {title}
  {\bibinfo {title} {Single-photon-level quantum memory at room temperature},\
  }\href {https://doi.org/10.1103/PhysRevLett.107.053603} {\bibfield  {journal}
  {\bibinfo  {journal} {Phys. Rev. Lett.}\ }\textbf {\bibinfo {volume} {107}},\
  \bibinfo {pages} {053603} (\bibinfo {year} {2011})}\BibitemShut {NoStop}%
\bibitem [{\citenamefont {Kaczmarek}\ \emph {et~al.}(2018)\citenamefont
  {Kaczmarek}, \citenamefont {Ledingham}, \citenamefont {Brecht}, \citenamefont
  {Thomas}, \citenamefont {Thekkadath}, \citenamefont {Lazo-Arjona},
  \citenamefont {Munns}, \citenamefont {Poem}, \citenamefont {Feizpour},
  \citenamefont {Saunders}, \citenamefont {Nunn},\ and\ \citenamefont
  {Walmsley}}]{Kaczmarek2018}%
  \BibitemOpen
  \bibfield  {author} {\bibinfo {author} {\bibfnamefont {K.~T.}\ \bibnamefont
  {Kaczmarek}}, \bibinfo {author} {\bibfnamefont {P.~M.}\ \bibnamefont
  {Ledingham}}, \bibinfo {author} {\bibfnamefont {B.}~\bibnamefont {Brecht}},
  \bibinfo {author} {\bibfnamefont {S.~E.}\ \bibnamefont {Thomas}}, \bibinfo
  {author} {\bibfnamefont {G.~S.}\ \bibnamefont {Thekkadath}}, \bibinfo
  {author} {\bibfnamefont {O.}~\bibnamefont {Lazo-Arjona}}, \bibinfo {author}
  {\bibfnamefont {J.~H.~D.}\ \bibnamefont {Munns}}, \bibinfo {author}
  {\bibfnamefont {E.}~\bibnamefont {Poem}}, \bibinfo {author} {\bibfnamefont
  {A.}~\bibnamefont {Feizpour}}, \bibinfo {author} {\bibfnamefont {D.~J.}\
  \bibnamefont {Saunders}}, \bibinfo {author} {\bibfnamefont {J.}~\bibnamefont
  {Nunn}},\ and\ \bibinfo {author} {\bibfnamefont {I.~A.}\ \bibnamefont
  {Walmsley}},\ }\bibfield  {title} {\bibinfo {title} {High-speed noise-free
  optical quantum memory},\ }\href {https://doi.org/10.1103/physreva.97.042316}
  {\bibfield  {journal} {\bibinfo  {journal} {Phys. Rev. A}\ }\textbf {\bibinfo
  {volume} {97}},\ \bibinfo {pages} {042316} (\bibinfo {year}
  {2018})}\BibitemShut {NoStop}%
\bibitem [{\citenamefont {Finkelstein}\ \emph {et~al.}(2018)\citenamefont
  {Finkelstein}, \citenamefont {Poem}, \citenamefont {Michel}, \citenamefont
  {Lahad},\ and\ \citenamefont {Firstenberg}}]{Finkelstein2018}%
  \BibitemOpen
  \bibfield  {author} {\bibinfo {author} {\bibfnamefont {R.}~\bibnamefont
  {Finkelstein}}, \bibinfo {author} {\bibfnamefont {E.}~\bibnamefont {Poem}},
  \bibinfo {author} {\bibfnamefont {O.}~\bibnamefont {Michel}}, \bibinfo
  {author} {\bibfnamefont {O.}~\bibnamefont {Lahad}},\ and\ \bibinfo {author}
  {\bibfnamefont {O.}~\bibnamefont {Firstenberg}},\ }\bibfield  {title}
  {\bibinfo {title} {Fast, noise-free memory for photon synchronization at room
  temperature},\ }\href {https://doi.org/10.1126/sciadv.aap8598} {\bibfield
  {journal} {\bibinfo  {journal} {Sci. Adv.}\ }\textbf {\bibinfo {volume}
  {4}},\ \bibinfo {pages} {aap8598} (\bibinfo {year} {2018})}\BibitemShut
  {NoStop}%
\bibitem [{\citenamefont {Pittman}\ \emph {et~al.}(2002)\citenamefont
  {Pittman}, \citenamefont {Jacobs},\ and\ \citenamefont
  {Franson}}]{Pittman2002}%
  \BibitemOpen
  \bibfield  {author} {\bibinfo {author} {\bibfnamefont {T.~B.}\ \bibnamefont
  {Pittman}}, \bibinfo {author} {\bibfnamefont {B.~C.}\ \bibnamefont
  {Jacobs}},\ and\ \bibinfo {author} {\bibfnamefont {J.~D.}\ \bibnamefont
  {Franson}},\ }\bibfield  {title} {\bibinfo {title} {Single photons on
  pseudodemand from stored parametric down-conversion},\ }\href
  {https://doi.org/10.1103/physreva.66.042303} {\bibfield  {journal} {\bibinfo
  {journal} {Phys. Rev. A}\ }\textbf {\bibinfo {volume} {66}},\ \bibinfo
  {pages} {042303} (\bibinfo {year} {2002})}\BibitemShut {NoStop}%
\bibitem [{\citenamefont {Migdall}\ \emph {et~al.}(2002)\citenamefont
  {Migdall}, \citenamefont {Branning},\ and\ \citenamefont
  {Castelletto}}]{Migdall2002}%
  \BibitemOpen
  \bibfield  {author} {\bibinfo {author} {\bibfnamefont {A.~L.}\ \bibnamefont
  {Migdall}}, \bibinfo {author} {\bibfnamefont {D.}~\bibnamefont {Branning}},\
  and\ \bibinfo {author} {\bibfnamefont {S.}~\bibnamefont {Castelletto}},\
  }\bibfield  {title} {\bibinfo {title} {Tailoring single-photon and
  multiphoton probabilities of a single-photon on-demand source},\ }\href
  {https://doi.org/10.1103/physreva.66.053805} {\bibfield  {journal} {\bibinfo
  {journal} {Phys. Rev. A}\ }\textbf {\bibinfo {volume} {66}},\ \bibinfo
  {pages} {053805} (\bibinfo {year} {2002})}\BibitemShut {NoStop}%
\bibitem [{\citenamefont {Kaneda}\ \emph {et~al.}(2015)\citenamefont {Kaneda},
  \citenamefont {Christensen}, \citenamefont {Wong}, \citenamefont {Park},
  \citenamefont {McCusker},\ and\ \citenamefont {Kwiat}}]{Kaneda2015}%
  \BibitemOpen
  \bibfield  {author} {\bibinfo {author} {\bibfnamefont {F.}~\bibnamefont
  {Kaneda}}, \bibinfo {author} {\bibfnamefont {B.~G.}\ \bibnamefont
  {Christensen}}, \bibinfo {author} {\bibfnamefont {J.~J.}\ \bibnamefont
  {Wong}}, \bibinfo {author} {\bibfnamefont {H.~S.}\ \bibnamefont {Park}},
  \bibinfo {author} {\bibfnamefont {K.~T.}\ \bibnamefont {McCusker}},\ and\
  \bibinfo {author} {\bibfnamefont {P.~G.}\ \bibnamefont {Kwiat}},\ }\bibfield
  {title} {\bibinfo {title} {Time-multiplexed heralded single-photon source},\
  }\href {https://doi.org/10.1364/optica.2.001010} {\bibfield  {journal}
  {\bibinfo  {journal} {Optica}\ }\textbf {\bibinfo {volume} {2}},\ \bibinfo
  {pages} {1010} (\bibinfo {year} {2015})}\BibitemShut {NoStop}%
\bibitem [{\citenamefont {Kaneda}\ and\ \citenamefont
  {Kwiat}(2019)}]{Kaneda2019}%
  \BibitemOpen
  \bibfield  {author} {\bibinfo {author} {\bibfnamefont {F.}~\bibnamefont
  {Kaneda}}\ and\ \bibinfo {author} {\bibfnamefont {P.~G.}\ \bibnamefont
  {Kwiat}},\ }\bibfield  {title} {\bibinfo {title} {High-efficiency
  single-photon generation via large-scale active time multiplexing},\ }\href
  {https://doi.org/10.1126/sciadv.aaw8586} {\bibfield  {journal} {\bibinfo
  {journal} {Sci. Adv.}\ }\textbf {\bibinfo {volume} {5}},\ \bibinfo {pages}
  {eaaw8586} (\bibinfo {year} {2019})}\BibitemShut {NoStop}%
\bibitem [{\citenamefont {Xiong}\ \emph {et~al.}(2016)\citenamefont {Xiong},
  \citenamefont {Zhang}, \citenamefont {Liu}, \citenamefont {Collins},
  \citenamefont {Mahendra}, \citenamefont {Helt}, \citenamefont {Steel},
  \citenamefont {Choi}, \citenamefont {Chae}, \citenamefont {Leong},\ and\
  \citenamefont {Eggleton}}]{Xiong2016}%
  \BibitemOpen
  \bibfield  {author} {\bibinfo {author} {\bibfnamefont {C.}~\bibnamefont
  {Xiong}}, \bibinfo {author} {\bibfnamefont {X.}~\bibnamefont {Zhang}},
  \bibinfo {author} {\bibfnamefont {Z.}~\bibnamefont {Liu}}, \bibinfo {author}
  {\bibfnamefont {M.~J.}\ \bibnamefont {Collins}}, \bibinfo {author}
  {\bibfnamefont {A.}~\bibnamefont {Mahendra}}, \bibinfo {author}
  {\bibfnamefont {L.~G.}\ \bibnamefont {Helt}}, \bibinfo {author}
  {\bibfnamefont {M.~J.}\ \bibnamefont {Steel}}, \bibinfo {author}
  {\bibfnamefont {D.~Y.}\ \bibnamefont {Choi}}, \bibinfo {author}
  {\bibfnamefont {C.~J.}\ \bibnamefont {Chae}}, \bibinfo {author}
  {\bibfnamefont {P.~H.~W.}\ \bibnamefont {Leong}},\ and\ \bibinfo {author}
  {\bibfnamefont {B.~J.}\ \bibnamefont {Eggleton}},\ }\bibfield  {title}
  {\bibinfo {title} {Active temporal multiplexing of indistinguishable heralded
  single photons},\ }\href {https://doi.org/10.1038/ncomms10853} {\bibfield
  {journal} {\bibinfo  {journal} {Nat. Commun.}\ }\textbf {\bibinfo {volume}
  {7}},\ \bibinfo {pages} {10853} (\bibinfo {year} {2016})}\BibitemShut
  {NoStop}%
\bibitem [{\citenamefont {Makino}\ \emph {et~al.}(2016)\citenamefont {Makino},
  \citenamefont {Hashimoto}, \citenamefont {ichi Yoshikawa}, \citenamefont
  {Ohdan}, \citenamefont {Toyama}, \citenamefont {van Loock},\ and\
  \citenamefont {Furusawa}}]{Makino2016}%
  \BibitemOpen
  \bibfield  {author} {\bibinfo {author} {\bibfnamefont {K.}~\bibnamefont
  {Makino}}, \bibinfo {author} {\bibfnamefont {Y.}~\bibnamefont {Hashimoto}},
  \bibinfo {author} {\bibfnamefont {J.}~\bibnamefont {Yoshikawa}},
  \bibinfo {author} {\bibfnamefont {H.}~\bibnamefont {Ohdan}}, \bibinfo
  {author} {\bibfnamefont {T.}~\bibnamefont {Toyama}}, \bibinfo {author}
  {\bibfnamefont {P.}~\bibnamefont {van Loock}},\ and\ \bibinfo {author}
  {\bibfnamefont {A.}~\bibnamefont {Furusawa}},\ }\bibfield  {title} {\bibinfo
  {title} {Synchronization of optical photons for quantum information
  processing},\ }\href {https://doi.org/10.1126/sciadv.1501772} {\bibfield
  {journal} {\bibinfo  {journal} {Sci. Adv.}\ }\textbf {\bibinfo {volume}
  {2}},\ \bibinfo {pages} {e1501772} (\bibinfo {year} {2016})}\BibitemShut
  {NoStop}%
\bibitem [{\citenamefont {Nunn}\ \emph {et~al.}(2013)\citenamefont {Nunn},
  \citenamefont {Langford}, \citenamefont {Kolthammer}, \citenamefont
  {Champion}, \citenamefont {Sprague}, \citenamefont {Michelberger},
  \citenamefont {Jin}, \citenamefont {England},\ and\ \citenamefont
  {Walmsley}}]{PhysRevLett.110.133601}%
  \BibitemOpen
  \bibfield  {author} {\bibinfo {author} {\bibfnamefont {J.}~\bibnamefont
  {Nunn}}, \bibinfo {author} {\bibfnamefont {N.~K.}\ \bibnamefont {Langford}},
  \bibinfo {author} {\bibfnamefont {W.~S.}\ \bibnamefont {Kolthammer}},
  \bibinfo {author} {\bibfnamefont {T.~F.~M.}\ \bibnamefont {Champion}},
  \bibinfo {author} {\bibfnamefont {M.~R.}\ \bibnamefont {Sprague}}, \bibinfo
  {author} {\bibfnamefont {P.~S.}\ \bibnamefont {Michelberger}}, \bibinfo
  {author} {\bibfnamefont {X.-M.}\ \bibnamefont {Jin}}, \bibinfo {author}
  {\bibfnamefont {D.~G.}\ \bibnamefont {England}},\ and\ \bibinfo {author}
  {\bibfnamefont {I.~A.}\ \bibnamefont {Walmsley}},\ }\bibfield  {title}
  {\bibinfo {title} {Enhancing multiphoton rates with quantum memories},\
  }\href {https://doi.org/10.1103/PhysRevLett.110.133601} {\bibfield  {journal}
  {\bibinfo  {journal} {Phys. Rev. Lett.}\ }\textbf {\bibinfo {volume} {110}},\
  \bibinfo {pages} {133601} (\bibinfo {year} {2013})}\BibitemShut {NoStop}%
\bibitem [{\citenamefont {Zhong}\ \emph {et~al.}(2020)\citenamefont {Zhong},
  \citenamefont {Wang}, \citenamefont {Deng}, \citenamefont {Chen},
  \citenamefont {Peng}, \citenamefont {Luo}, \citenamefont {Qin}, \citenamefont
  {Wu}, \citenamefont {Ding}, \citenamefont {Hu}, \citenamefont {Hu},
  \citenamefont {Yang}, \citenamefont {Zhang}, \citenamefont {Li},
  \citenamefont {Li}, \citenamefont {Jiang}, \citenamefont {Gan}, \citenamefont
  {Yang}, \citenamefont {You}, \citenamefont {Wang}, \citenamefont {Li},
  \citenamefont {Liu}, \citenamefont {Lu},\ and\ \citenamefont
  {Pan}}]{ZhongScience2020}%
  \BibitemOpen
  \bibfield  {author} {\bibinfo {author} {\bibfnamefont {H.-S.}\ \bibnamefont
  {Zhong}}, \bibinfo {author} {\bibfnamefont {H.}~\bibnamefont {Wang}},
  \bibinfo {author} {\bibfnamefont {Y.-H.}\ \bibnamefont {Deng}}, \bibinfo
  {author} {\bibfnamefont {M.-C.}\ \bibnamefont {Chen}}, \bibinfo {author}
  {\bibfnamefont {L.-C.}\ \bibnamefont {Peng}}, \bibinfo {author}
  {\bibfnamefont {Y.-H.}\ \bibnamefont {Luo}}, \bibinfo {author} {\bibfnamefont
  {J.}~\bibnamefont {Qin}}, \bibinfo {author} {\bibfnamefont {D.}~\bibnamefont
  {Wu}}, \bibinfo {author} {\bibfnamefont {X.}~\bibnamefont {Ding}}, \bibinfo
  {author} {\bibfnamefont {Y.}~\bibnamefont {Hu}}, \bibinfo {author}
  {\bibfnamefont {P.}~\bibnamefont {Hu}}, \bibinfo {author} {\bibfnamefont
  {X.-Y.}\ \bibnamefont {Yang}}, \bibinfo {author} {\bibfnamefont {W.-J.}\
  \bibnamefont {Zhang}}, \bibinfo {author} {\bibfnamefont {H.}~\bibnamefont
  {Li}}, \bibinfo {author} {\bibfnamefont {Y.}~\bibnamefont {Li}}, \bibinfo
  {author} {\bibfnamefont {X.}~\bibnamefont {Jiang}}, \bibinfo {author}
  {\bibfnamefont {L.}~\bibnamefont {Gan}}, \bibinfo {author} {\bibfnamefont
  {G.}~\bibnamefont {Yang}}, \bibinfo {author} {\bibfnamefont {L.}~\bibnamefont
  {You}}, \bibinfo {author} {\bibfnamefont {Z.}~\bibnamefont {Wang}}, \bibinfo
  {author} {\bibfnamefont {L.}~\bibnamefont {Li}}, \bibinfo {author}
  {\bibfnamefont {N.-L.}\ \bibnamefont {Liu}}, \bibinfo {author} {\bibfnamefont
  {C.-Y.}\ \bibnamefont {Lu}},\ and\ \bibinfo {author} {\bibfnamefont {J.-W.}\
  \bibnamefont {Pan}},\ }\bibfield  {title} {\bibinfo {title} {Quantum
  computational advantage using photons},\ }\href
  {https://doi.org/10.1126/science.abe8770} {\bibfield  {journal} {\bibinfo
  {journal} {Science}\ }\textbf {\bibinfo {volume} {370}},\ \bibinfo {pages}
  {1460} (\bibinfo {year} {2020})}\ \BibitemShut {NoStop}%
\bibitem [{\citenamefont {de~Sterke}\ \emph {et~al.}(2002)\citenamefont
  {de~Sterke}, \citenamefont {Tsoy},\ and\ \citenamefont {Sipe}}]{Sterke2002}%
  \BibitemOpen
  \bibfield  {author} {\bibinfo {author} {\bibfnamefont {C.~M.}\ \bibnamefont
  {de~Sterke}}, \bibinfo {author} {\bibfnamefont {E.~N.}\ \bibnamefont
  {Tsoy}},\ and\ \bibinfo {author} {\bibfnamefont {J.~E.}\ \bibnamefont
  {Sipe}},\ }\bibfield  {title} {\bibinfo {title} {Light trapping in a fiber
  grating defect by four-wave mixing},\ }\href
  {https://doi.org/10.1364/ol.27.000485} {\bibfield  {journal} {\bibinfo
  {journal} {Opt. Lett.}\ }\textbf {\bibinfo {volume} {27}},\ \bibinfo {pages}
  {485} (\bibinfo {year} {2002})}\BibitemShut {NoStop}%
\bibitem [{\citenamefont {Leung}\ and\ \citenamefont
  {Ralph}(2006)}]{Leung2006}%
  \BibitemOpen
  \bibfield  {author} {\bibinfo {author} {\bibfnamefont {P.~M.}\ \bibnamefont
  {Leung}}\ and\ \bibinfo {author} {\bibfnamefont {T.~C.}\ \bibnamefont
  {Ralph}},\ }\bibfield  {title} {\bibinfo {title} {Quantum memory scheme based
  on optical fibers and cavities},\ }\href
  {https://doi.org/10.1103/physreva.74.022311} {\bibfield  {journal} {\bibinfo
  {journal} {Phys. Rev. A}\ }\textbf {\bibinfo {volume} {74}},\ \bibinfo
  {pages} {022311} (\bibinfo {year} {2006})}\BibitemShut {NoStop}%
\bibitem [{\citenamefont {Margulis}\ \emph {et~al.}(2011)\citenamefont
  {Margulis}, \citenamefont {Yu}, \citenamefont {Malmstr\"{o}m}, \citenamefont
  {Rugeland}, \citenamefont {Knape},\ and\ \citenamefont
  {Tarasenko}}]{Margulis2011}%
  \BibitemOpen
  \bibfield  {author} {\bibinfo {author} {\bibfnamefont {W.}~\bibnamefont
  {Margulis}}, \bibinfo {author} {\bibfnamefont {Z.}~\bibnamefont {Yu}},
  \bibinfo {author} {\bibfnamefont {M.}~\bibnamefont {Malmstr\"{o}m}}, \bibinfo
  {author} {\bibfnamefont {P.}~\bibnamefont {Rugeland}}, \bibinfo {author}
  {\bibfnamefont {H.}~\bibnamefont {Knape}},\ and\ \bibinfo {author}
  {\bibfnamefont {O.}~\bibnamefont {Tarasenko}},\ }\bibfield  {title} {\bibinfo
  {title} {High-speed electrical switching in optical fibers [invited]},\
  }\href {https://doi.org/10.1364/ao.50.000e65} {\bibfield  {journal} {\bibinfo
   {journal} {Appl. Opt.}\ }\textbf {\bibinfo {volume} {50}},\ \bibinfo {pages}
  {E65} (\bibinfo {year} {2011})}\BibitemShut {NoStop}%
\bibitem [{\citenamefont {McKinstrie}\ \emph {et~al.}(2005)\citenamefont
  {McKinstrie}, \citenamefont {Harvey}, \citenamefont {Radic},\ and\
  \citenamefont {Raymer}}]{OptExpress.13.9131}%
  \BibitemOpen
  \bibfield  {author} {\bibinfo {author} {\bibfnamefont {C.~J.}\ \bibnamefont
  {McKinstrie}}, \bibinfo {author} {\bibfnamefont {J.~D.}\ \bibnamefont
  {Harvey}}, \bibinfo {author} {\bibfnamefont {S.}~\bibnamefont {Radic}},\ and\
  \bibinfo {author} {\bibfnamefont {M.~G.}\ \bibnamefont {Raymer}},\ }\bibfield
   {title} {\bibinfo {title} {Translation of quantum states by four-wave mixing
  in fibers},\ }\href {https://doi.org/10.1364/OPEX.13.009131} {\bibfield
  {journal} {\bibinfo  {journal} {Opt. Express}\ }\textbf {\bibinfo {volume}
  {13}},\ \bibinfo {pages} {9131} (\bibinfo {year} {2005})}\BibitemShut
  {NoStop}%
\bibitem [{\citenamefont {Kupchak}\ \emph {et~al.}(2019)\citenamefont
  {Kupchak}, \citenamefont {Erskine}, \citenamefont {England},\ and\
  \citenamefont {Sussman}}]{Kupchak2019}%
  \BibitemOpen
  \bibfield  {author} {\bibinfo {author} {\bibfnamefont {C.}~\bibnamefont
  {Kupchak}}, \bibinfo {author} {\bibfnamefont {J.}~\bibnamefont {Erskine}},
  \bibinfo {author} {\bibfnamefont {D.}~\bibnamefont {England}},\ and\ \bibinfo
  {author} {\bibfnamefont {B.}~\bibnamefont {Sussman}},\ }\bibfield  {title}
  {\bibinfo {title} {Terahertz-bandwidth switching of heralded single
  photons},\ }\href {https://doi.org/10.1364/ol.44.001427} {\bibfield
  {journal} {\bibinfo  {journal} {Opt. Lett.}\ }\textbf {\bibinfo {volume}
  {44}},\ \bibinfo {pages} {1427} (\bibinfo {year} {2019})}\BibitemShut
  {NoStop}%
\bibitem [{\citenamefont {Bouchard}\ \emph {et~al.}(2021)\citenamefont
  {Bouchard}, \citenamefont {England}, \citenamefont {Bustard}, \citenamefont
  {Fenwick}, \citenamefont {Karimi}, \citenamefont {Heshami},\ and\
  \citenamefont {Sussman}}]{Bouchard2021}%
  \BibitemOpen
  \bibfield  {author} {\bibinfo {author} {\bibfnamefont {F.}~\bibnamefont
  {Bouchard}}, \bibinfo {author} {\bibfnamefont {D.}~\bibnamefont {England}},
  \bibinfo {author} {\bibfnamefont {P.~J.}\ \bibnamefont {Bustard}}, \bibinfo
  {author} {\bibfnamefont {K.~L.}\ \bibnamefont {Fenwick}}, \bibinfo {author}
  {\bibfnamefont {E.}~\bibnamefont {Karimi}}, \bibinfo {author} {\bibfnamefont
  {K.}~\bibnamefont {Heshami}},\ and\ \bibinfo {author} {\bibfnamefont
  {B.}~\bibnamefont {Sussman}},\ }\bibfield  {title} {\bibinfo {title}
  {Achieving ultimate noise tolerance in quantum communication},\ }\href
  {https://doi.org/10.1103/physrevapplied.15.024027} {\bibfield  {journal}
  {\bibinfo  {journal} {Phys. Rev. Appl}\ }\textbf {\bibinfo {volume} {15}},\
  \bibinfo {pages} {024027} (\bibinfo {year} {2021})}\BibitemShut {NoStop}%
\bibitem [{\citenamefont {McGuinness}\ \emph {et~al.}(2010)\citenamefont
  {McGuinness}, \citenamefont {Raymer}, \citenamefont {McKinstrie},\ and\
  \citenamefont {Radic}}]{McGuinness2010}%
  \BibitemOpen
  \bibfield  {author} {\bibinfo {author} {\bibfnamefont {H.~J.}\ \bibnamefont
  {McGuinness}}, \bibinfo {author} {\bibfnamefont {M.~G.}\ \bibnamefont
  {Raymer}}, \bibinfo {author} {\bibfnamefont {C.~J.}\ \bibnamefont
  {McKinstrie}},\ and\ \bibinfo {author} {\bibfnamefont {S.}~\bibnamefont
  {Radic}},\ }\bibfield  {title} {\bibinfo {title} {Quantum frequency
  translation of single-photon states in a photonic crystal fiber},\ }\href
  {https://doi.org/10.1103/PhysRevLett.105.093604} {\bibfield  {journal}
  {\bibinfo  {journal} {Phys. Rev. Lett.}\ }\textbf {\bibinfo {volume} {105}},\
  \bibinfo {pages} {093604} (\bibinfo {year} {2010})}\BibitemShut {NoStop}%
\bibitem [{\citenamefont {Clark}\ \emph {et~al.}(2013)\citenamefont {Clark},
  \citenamefont {Shahnia}, \citenamefont {Collins}, \citenamefont {Xiong},\
  and\ \citenamefont {Eggleton}}]{OptLett.38.947}%
  \BibitemOpen
  \bibfield  {author} {\bibinfo {author} {\bibfnamefont {A.~S.}\ \bibnamefont
  {Clark}}, \bibinfo {author} {\bibfnamefont {S.}~\bibnamefont {Shahnia}},
  \bibinfo {author} {\bibfnamefont {M.~J.}\ \bibnamefont {Collins}}, \bibinfo
  {author} {\bibfnamefont {C.}~\bibnamefont {Xiong}},\ and\ \bibinfo {author}
  {\bibfnamefont {B.~J.}\ \bibnamefont {Eggleton}},\ }\bibfield  {title}
  {\bibinfo {title} {High-efficiency frequency conversion in the single-photon
  regime},\ }\href {https://doi.org/10.1364/OL.38.000947} {\bibfield  {journal}
  {\bibinfo  {journal} {Opt. Lett.}\ }\textbf {\bibinfo {volume} {38}},\
  \bibinfo {pages} {947} (\bibinfo {year} {2013})}\BibitemShut {NoStop}%
\bibitem [{\citenamefont {Clemmen}\ \emph {et~al.}(2018)\citenamefont
  {Clemmen}, \citenamefont {Farsi}, \citenamefont {Ramelow},\ and\
  \citenamefont {Gaeta}}]{Clemmen2018}%
  \BibitemOpen
  \bibfield  {author} {\bibinfo {author} {\bibfnamefont {S.}~\bibnamefont
  {Clemmen}}, \bibinfo {author} {\bibfnamefont {A.}~\bibnamefont {Farsi}},
  \bibinfo {author} {\bibfnamefont {S.}~\bibnamefont {Ramelow}},\ and\ \bibinfo
  {author} {\bibfnamefont {A.~L.}\ \bibnamefont {Gaeta}},\ }\bibfield  {title}
  {\bibinfo {title} {All-optically tunable buffer for single photons},\ }\href
  {https://doi.org/10.1364/ol.43.002138} {\bibfield  {journal} {\bibinfo
  {journal} {Opt. Lett.}\ }\textbf {\bibinfo {volume} {43}},\ \bibinfo {pages}
  {2138} (\bibinfo {year} {2018})}\BibitemShut {NoStop}%
\bibitem [{\citenamefont {Smith}\ \emph {et~al.}(2009)\citenamefont {Smith},
  \citenamefont {Mahou}, \citenamefont {Cohen}, \citenamefont {Lundeen},\ and\
  \citenamefont {Walmsley}}]{OptExpress.17.23589}%
  \BibitemOpen
  \bibfield  {author} {\bibinfo {author} {\bibfnamefont {B.~J.}\ \bibnamefont
  {Smith}}, \bibinfo {author} {\bibfnamefont {P.}~\bibnamefont {Mahou}},
  \bibinfo {author} {\bibfnamefont {O.}~\bibnamefont {Cohen}}, \bibinfo
  {author} {\bibfnamefont {J.~S.}\ \bibnamefont {Lundeen}},\ and\ \bibinfo
  {author} {\bibfnamefont {I.~A.}\ \bibnamefont {Walmsley}},\ }\bibfield
  {title} {\bibinfo {title} {Photon pair generation in birefringent optical
  fibers},\ }\href {https://doi.org/10.1364/OE.17.023589} {\bibfield  {journal}
  {\bibinfo  {journal} {Opt. Express}\ }\textbf {\bibinfo {volume} {17}},\
  \bibinfo {pages} {23589} (\bibinfo {year} {2009})}\BibitemShut {NoStop}%
\bibitem [{\citenamefont {Christensen}\ \emph {et~al.}(2018)\citenamefont
  {Christensen}, \citenamefont {Koefoed}, \citenamefont {Bell}, \citenamefont
  {McKinstrie},\ and\ \citenamefont {Rottwitt}}]{OptExpress.26.17145}%
  \BibitemOpen
  \bibfield  {author} {\bibinfo {author} {\bibfnamefont {J.~B.}\ \bibnamefont
  {Christensen}}, \bibinfo {author} {\bibfnamefont {J.~G.}\ \bibnamefont
  {Koefoed}}, \bibinfo {author} {\bibfnamefont {B.~A.}\ \bibnamefont {Bell}},
  \bibinfo {author} {\bibfnamefont {C.~J.}\ \bibnamefont {McKinstrie}},\ and\
  \bibinfo {author} {\bibfnamefont {K.}~\bibnamefont {Rottwitt}},\ }\bibfield
  {title} {\bibinfo {title} {Shape-preserving and unidirectional frequency
  conversion by four-wave mixing},\ }\href
  {https://doi.org/10.1364/OE.26.017145} {\bibfield  {journal} {\bibinfo
  {journal} {Opt. Express}\ }\textbf {\bibinfo {volume} {26}},\ \bibinfo
  {pages} {17145} (\bibinfo {year} {2018})}\BibitemShut {NoStop}%
\bibitem [{\citenamefont {Weiner}\ \emph {et~al.}(1988)\citenamefont {Weiner},
  \citenamefont {Heritage},\ and\ \citenamefont {Kirschner}}]{Weiner1988}%
  \BibitemOpen
  \bibfield  {author} {\bibinfo {author} {\bibfnamefont {A.~M.}\ \bibnamefont
  {Weiner}}, \bibinfo {author} {\bibfnamefont {J.~P.}\ \bibnamefont
  {Heritage}},\ and\ \bibinfo {author} {\bibfnamefont {E.~M.}\ \bibnamefont
  {Kirschner}},\ }\bibfield  {title} {\bibinfo {title} {High-resolution
  femtosecond pulse shaping},\ }\href {https://doi.org/10.1364/josab.5.001563}
  {\bibfield  {journal} {\bibinfo  {journal} {J. Opt. Soc. Am. B}\ }\textbf
  {\bibinfo {volume} {5}},\ \bibinfo {pages} {1563} (\bibinfo {year}
  {1988})}\BibitemShut {NoStop}%
\bibitem [{Note1()}]{Note1}%
  \BibitemOpen
  \bibinfo {note} {See Supplemental Material for experimental
  details}\BibitemShut {NoStop}%
\bibitem [{Note2()}]{Note2}%
  \BibitemOpen
  \bibinfo {note} {See Supplemental Material for noise
  measurements.}\BibitemShut {Stop}%
\bibitem [{\citenamefont {Stolen}\ \emph {et~al.}(1989)\citenamefont {Stolen},
  \citenamefont {Gordon}, \citenamefont {Tomlinson},\ and\ \citenamefont
  {Haus}}]{JOSAB.6.1159}%
  \BibitemOpen
  \bibfield  {author} {\bibinfo {author} {\bibfnamefont {R.~H.}\ \bibnamefont
  {Stolen}}, \bibinfo {author} {\bibfnamefont {J.~P.}\ \bibnamefont {Gordon}},
  \bibinfo {author} {\bibfnamefont {W.~J.}\ \bibnamefont {Tomlinson}},\ and\
  \bibinfo {author} {\bibfnamefont {H.~A.}\ \bibnamefont {Haus}},\ }\bibfield
  {title} {\bibinfo {title} {Raman response function of silica-core fibers},\
  }\href {https://doi.org/10.1364/JOSAB.6.001159} {\bibfield  {journal}
  {\bibinfo  {journal} {J. Opt. Soc. Am. B}\ }\textbf {\bibinfo {volume} {6}},\
  \bibinfo {pages} {1159} (\bibinfo {year} {1989})}\BibitemShut {NoStop}%
\end{thebibliography}
\end{document}


\preprint{APS/123-QED}

\title{Toward a Quantum Memory in a Fiber Cavity Controlled by Intracavity Frequency Translation\\Supplemental Material}

\author{Philip~J. Bustard}
 \email{philip.bustard@nrc-cnrc.gc.ca}
\affiliation{National Research Council of Canada, 100 Sussex Drive, Ottawa, Ontario, K1A 0R6, Canada}
\author{Kent Bonsma-Fisher}
\affiliation{National Research Council of Canada, 100 Sussex Drive, Ottawa, Ontario, K1A 0R6, Canada}
\author{Cyril Hnatovsky}
\affiliation{National Research Council of Canada, 100 Sussex Drive, Ottawa, Ontario, K1A 0R6, Canada}
\author{Dan Grobnic}
\affiliation{National Research Council of Canada, 100 Sussex Drive, Ottawa, Ontario, K1A 0R6, Canada}
\author{Stephen J. Mihailov}
\affiliation{National Research Council of Canada, 100 Sussex Drive, Ottawa, Ontario, K1A 0R6, Canada}
\author{Duncan England}
 \email{duncan.england@nrc-cnrc.gc.ca}
\affiliation{National Research Council of Canada, 100 Sussex Drive, Ottawa, Ontario, K1A 0R6, Canada}
\author{Benjamin~J. Sussman}
\affiliation{National Research Council of Canada, 100 Sussex Drive, Ottawa, Ontario, K1A 0R6, Canada}
\affiliation{Department of Physics, University of Ottawa, Ottawa, Ontario, K1N 6N5, Canada}
\date{\today}

\maketitle

\section{Photon detection efficiency and the single photon level}
When working with attenuated signal pulses, we set the single photon level by correcting for the total detection efficiency for signal photons at $\lambda_{\text{s}}=\unit[902.5]{nm}$ from inside the cavity. The total detection efficiency is $\eta_{\text{det}}=0.21(2)$, where $\eta_{\text{det}}=\mathcal{T}_{\text{s}}\eta_{\text{c}}\eta_{\text{SPCM}}$ for photons exiting the fibre. Here, $\mathcal{T}_{\text{s}}=0.74(5)$ is the end-facet-coating transmission at the output signal wavelength, $\eta_{\text{c}}=0.65(2)$ is the collection efficiency from the fiber-cavity exit facet to the SPCM, and $\eta_{\text{SPCM}}=0.44$ is the SPCM quantum efficiency quoted by the manufacturer. The quoted loss due to propagation in the fibre is \unit[5]{dB/km}; this is negligible during a single pass through the fibre.

\section{Power scaling of noise and Bragg-scattering four-wave mixing}
\begin{figure}
\begin{center}
\includegraphics[trim=115 113 158 462,clip,width=\columnwidth]{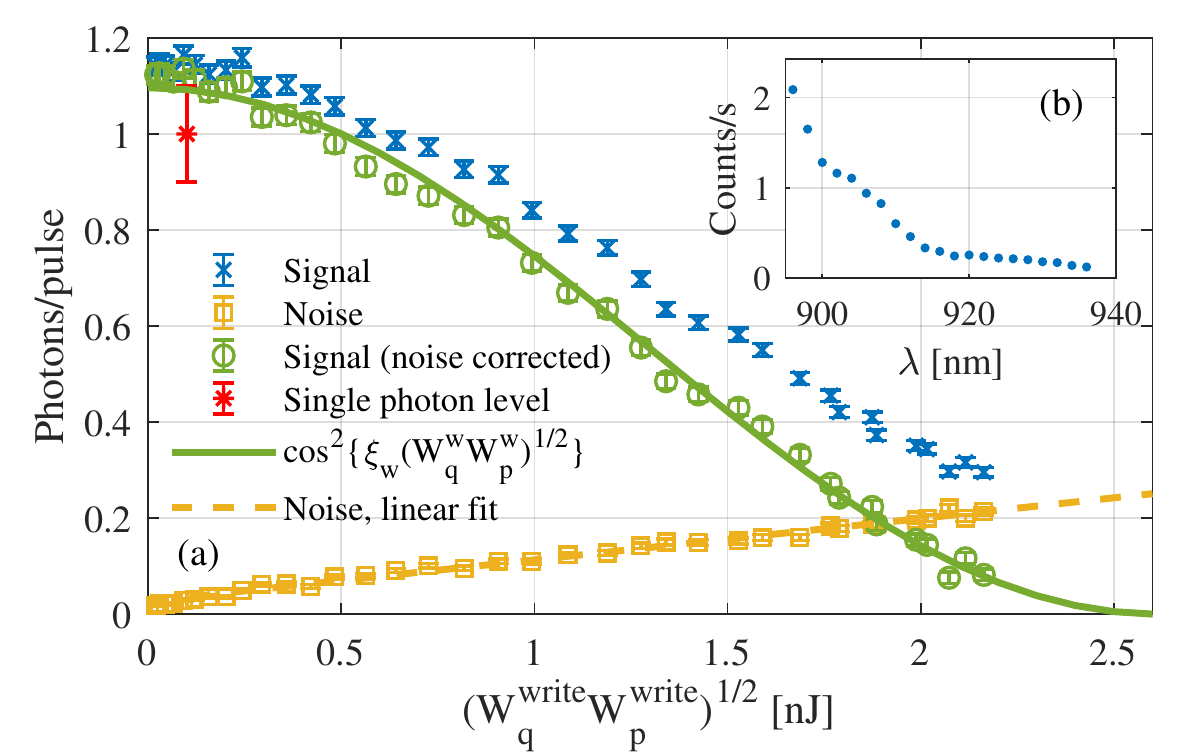}
\end{center}
\caption{(a) Signal (blue crosses), noise (yellow squares), and noise-corrected signal (green circles) as a function of control pulse energies for the write interaction. Fits to the noise-corrected signal (green, solid curve) and noise (yellow, dashed line) are plotted (see text). (b) Plot of the fast axis (Raman) noise spectrum in an uncoated HB800 PM fiber. \label{fig:writePowerScanSPL}}
\end{figure}

The input and output signal modes of Bragg-scattering four-wave mixing (BSFWM) in a birefringent fiber are related by a beam splitter formalism: the output signal pulse energy from the write interaction $W_{\text{s}}^{\text{out}}$ is expected to vary as 
\[
W_{\text{s}}^{\text{out}}=W_{\text{s}}^{\text{in}}\cos^2\left\{\xi_{\text{w}}\sqrt{W_{\text{q}}^{\text{w}}W_{\text{p}}^{\text{w}}}\right\},
\]
where $\xi_{\text{w}}$ depends on the pulse shapes and fiber parameters~\cite{OptExpress.26.17145}, $W_{\text{s}}^{\text{in}}$ is the input signal pulse energy, and $W_{\text{p,q}}^{\text{w}}$ are the write pulse energies in the fiber. Figure~\ref{fig:writePowerScanSPL}(a) is a plot of the signal (blue crosses), noise (yellow squares), noise-corrected signal (green circles), and a fit of $W_{\text{s}}^{\text{out}}$ (green, solid line) as a function of $\sqrt{W_{\text{q}}^{\text{w}}W_{\text{p}}^{\text{w}}}$.
The fit to the noise corrected signal has good qualitative agreement and $R^2=0.996$. A weighted linear regression on the noise data gives a good fit with $R^2>0.98$. To gain further understanding of the noise process, we exchanged the EFC fiber for an uncoated fiber of similar length and measured the noise spectrum from the write control pulses. Figure~\ref{fig:writePowerScanSPL}(b) shows the broadband noise spectrum gradually decreasing with increasing wavelength. This spectrum and the linear power dependence of the noise are consistent with spontaneous Raman scattering from the control pulses as the dominant source of noise.